\documentclass[galaxies,article,accept,pdftex,moreauthors]{Definitions/mdpi} 
\usepackage{multirow}
\usepackage{longtable}

\firstpage{1} 
\pubvolume{1}
\issuenum{1}
\articlenumber{0}
\pubyear{2024}
\copyrightyear{2024}
\externaleditor{Academic Editor: Firstname \linebreak~Lastname}
\datereceived{20 January 2024} 
\daterevised{12 February 2024} 
\dateaccepted{14 February 2024} 
\datepublished{ } 
\hreflink{https://doi.org/} 

\usepackage{natbib}

\Title{Superluminal Motion and Jet Parameters in the Gamma-ray-Emitting Narrow-Line Seyfert 1 Galaxy TXS~1206+549} 

\TitleCitation{Superluminal Motion and Jet Parameters in the Gamma-ray-Emitting Narrow-Line Seyfert 1 Galaxy TXS~1206+549}

\Author{Bettina Koz\'ak $^{1}$, S\'andor Frey $^{2,3,4}$*\orcidA{} and Krisztina \'Eva Gab\'anyi $^{1,2,3,5}$\orcidB{}}

\AuthorNames{Bettina Koz\'ak, S\'andor Frey and Krisztina \'Eva Gab\'anyi}

\AuthorCitation{Koz\'ak, B.; Frey, S.; Gab\'anyi,~K.\'E.}

\address{%
$^{1}$ \quad Department of Astronomy, Institute of Physics and Astronomy, ELTE---E\"otv\"os Lor\'and University, P\'azm\'any~P\'eter s\'et\'any 1/A, H-1117~Budapest, Hungary; k.betti0317@gmail.com~(B.K.); k.gabanyi@astro.elte.hu~(K.\'E.G.)\\
$^{2}$ \quad Konkoly Observatory, HUN-REN Research Centre for Astronomy and Earth Sciences (CSFK), Konkoly~Thege~Mikl\'os~\'ut~15-17, H-1121~Budapest, Hungary\\
$^{3}$ \quad CSFK, MTA Centre of Excellence, Konkoly Thege Mikl\'os \'ut 15-17, H-1121~Budapest, Hungary\\
$^{4}$ \quad Institute of Physics and Astronomy, ELTE---E\"otv\"os Lor\'and University, P\'azm\'any P\'eter s\'et\'any 1/A, H-1117~Budapest, Hungary\\
$^{5}$ \quad HUN-REN--ELTE Extragalactic Astrophysics Research Group, ELTE---E\"otv\"os Lor\'and University, P\'azm\'any~P\'eter s\'et\'any 1/A, H-1117~Budapest, Hungary}

\corres{Correspondence: frey.sandor@csfk.org}

\abstract{Narrow-line Seyfert 1 (NLS1) galaxies are a peculiar subclass of active galactic nuclei (AGN). Among them, TXS~1206+549 belongs to a small group of radio-loud and $\gamma$-ray-emitting NLS1 galaxies. We focus on the radio properties of this galaxy by analysing archival, high-resolution, very long baseline interferometry (VLBI) imaging observations taken at $8$~GHz frequency in six epochs between $1994$ and $2018$. Using the milliarcsecond-scale radio structure, we can resolve a core and a jet component whose angular separation increases by $(0.055 \pm 0.006)$~mas\,yr$^{-1}$. This corresponds to an apparent superluminal jet component motion of $(3.5 \pm 0.4)\,c$. From the core brightness temperature and the jet component proper motion, we determine the characteristic Doppler-boosting factor, the bulk Lorentz factor, and the jet viewing angle. We find no compelling evidence for a very closely aligned blazar-type jet. The parameters for TXS~1206$+$549 resemble those of radio-loud quasar jets with a moderate Lorentz factor ($\Gamma \approx 4$) and $\vartheta \approx 24^{\circ}$ inclination to the line of sight.}


\begin{document}

\section{Introduction}
\label{Intro}

Seyfert galaxies \cite{1943ApJ....97...28S} have bright, unresolved nuclei in optical images, with broad emission lines in their spectra. They typically reside in spiral host galaxies and belong to the general class of active galactic nuclei (AGN). These objects are divided into two main categories: Seyfert 1 (Sy1) and Seyfert 2 (Sy2) galaxies. According to the unified scheme of AGN, e.g., \cite{1993ARA&A..31..473A}, Sy1 galaxies are seen nearly face-on and the broad-line region (BLR), composed of gas clouds orbiting around the central supermassive black hole (SMBH) with speeds of several thousand km\,s$^{-1}$, is visible from our line of sight. This gives rise to permitted lines in the H\textsc{i} Balmer series that are broader than the forbidden lines \cite{1985ApJ...297..166O}. In turn, Sy2 galaxies are viewed nearly edge-on and the dusty torus obscures the radiation from the BLR. Therefore, Sy2 spectra are characterised by narrower emission lines coming from the so-called narrow-line region (NLR), i.e., from ionised gas clouds orbiting further away from the SMBH. The full width at half maximum (FWHM) of the forbidden lines is typically 300--800~km\,s$^{-1}$ for both Sy1 and Sy2, while the Balmer lines are much broader, reaching up to $6000$~km\,s$^{-1}$ in Sy1 galaxies \cite{1985ApJ...297..166O}. There are also Seyfert galaxies with intermediate properties~\cite{1983ApJ...273..489C}.  

There are also AGN with general Sy1 properties but a much narrower H$\beta$ emission line in their spectra compared to what is typical for the Sy1 group. These are called narrow-line Sy1 (NLS1) galaxies \cite{1985ApJ...297..166O}. Their defining characteristics are the H$\beta$ line with \mbox{FWHM $<~2000$~km\,s$^{-1}$}, and the relative weakness of the [O\textsc{iii}]$\lambda$5007~\AA\ emission, \mbox{[O\textsc{iii}]/H$\beta$ $< 3$ \cite{1989ApJ...342..224G}}. Because they usually exhibit strong Fe\textsc{ii} emissions, the presence of these lines in the spectrum is also often included in the definition.

The majority of AGN are radio-quiet or radio-silent; only a few  of them ($\sim$10\%) are radio-loud \cite{1995PASP..107..803U}. Because their radio emissions are dominated by the synchrotron emission of relativistic plasma jets ejected from the vicinity of the central SMBH, these are also called jetted AGN \cite{2017NatAs...1E.194P}. 
Almost all extragalactic $\gamma$-ray sources identified with counterparts in other electromagnetic wavebands in the 4th \textit{Fermi} Large-Area Telescope (LAT) catalogue are radio-loud jetted AGN \cite{2020ApJS..247...33A}.
A similar fraction of radio-loud vs. radio-quiet sources is found for the NLS1 sub-class: only about $\sim$8\% have a radio-loudness parameter, i.e., the radio to optical flux density ratio, $R = f_{\nu} \mathrm{(5\,GHz)}/ f_{\nu}\,(4400\,\text{\AA}) > 10$, and very few have \mbox{$R > 100$ \cite{2006AJ....132..531K}}. The detection of $\gamma$-ray emissions from four radio-loud NLS1 (RL-NLS1) objects after the first year of operations of the \textit{Fermi} LAT instrument \cite{2009ApJ...699..976A,2009ApJ...707L.142A} confirmed the existence of powerful relativistic jets in these AGN.

The observations suggest that radio-loud, jetted AGN are hosted in elliptical galaxies with high black hole (BH) mass. It seems that lower-mass ($M \lesssim 10^{8}$\,M$_{\odot}$) central BHs in late-type galaxies are generally not capable of launching powerful jets, establishing a strong connection between BH mass and radio loudness \cite{2000ApJ...543L.111L}. However, with the discovery of not just NLS1s but other types of radio galaxies \cite{1999ApJS..123...79H} exhibiting $\gamma$-ray emissions, this BH mass--radio loudness correlation has been under debate. NLS1s hosted in diverse types of galaxies, including spirals \cite{2020Univ....6..136F}, are considered to have low-mass ($\sim$10$^{6}$--10$^8\,M_{\odot}$) black holes and high accretion rates, close to or even above the Eddington limit  \cite{2002ApJ...565...78B,2020Univ....6..136F}. It is possible that the $\gamma$-ray-emitting RL-NLS1s occupy the upper end of the NLS1 mass spectrum \cite{2015A&A...573A..76J}. It has also been proposed that these are the low-luminosity counterparts of the flat-spectrum radio quasars (FSRQs), and there might be an evolutionary connection between them, as RL-NLS1 sources comprise the still-evolving, younger phases of FSRQs \cite{2016A&A...591A..98B}. There is no doubt that these peculiar $\gamma$-ray-emitting RL-NLS1s play an important role in understanding how jetted AGN differ from their radio-quiet counterparts and what contributes to the production of powerful \mbox{jets \cite{2019Galax...7...87D,2019JApA...40...39P}}. 

One of the few $\gamma$-ray-detected RL-NLS1s that was recently found \cite{2021MNRAS.504L..22R} is TXS~1206$+$549 (also known as J1208+5441) at a redshift of $z=1.344$ \cite{2005AJ....130..367S}. This is the most distant $\gamma$-ray NLS1 found to date. In the \textit{Fermi}-4LAC catalogue, it is classified as FSRQ \cite{2020ApJS..248...27T}.
The discoverers obtained a near-infrared spectrum and measured the FMWH of the H$\beta$ line as $(1194~\pm~77)$~km\,s$^{-1}$ and the [O\textsc{iii}] to H$\beta$ flux ratio as $\sim$0.7, so TXS~1206$+$549 satisfies the NLS1 criteria. Although an ultraviolet Fe\textsc{ii} line was found for the object, optical Fe\textsc{ii} lines were not seen in the \mbox{spectrum \cite{2021MNRAS.504L..22R}}. To estimate the BH mas in TXS~1206$+$549, different methods were employed,
resulting in values of $\sim$10$^{7}$--10$^{8}$~M$_{\odot}$, consistent with what is found for NLS1 objects in general and $\gamma$-ray-detected NLS1s in particular. These values are one or two orders of magnitude lower than the SMBH masses of powerful jetted AGN \cite{2021MNRAS.504L..22R}. The high radio loudness parameter $R \approx 1300$ indicates that TXS~1206$+$549 belongs to RL-NLS1s, although the variability known in both the optical and radio wavebands means that $R$ is also variable.   

Most recently, the presence of a quasiperiodic oscillation (QPO) with a period of about 134~d in the Zwicky Transient Facility optical light curve of TXS~1206+549 was proposed \cite{2024ApJ...961...38R}. The signal was present between 2018 and 2021. QPOs are usually found at multiple wavebands in AGN, including NLS1 sources, but their physical origin is not well understood a present.

\textls[-15]{TXS~1206$+$549 is a well-known, flat-spectrum radio source, with variable (\mbox{$\sim$120--400~mJy})} total flux density across two orders of magnitude in frequency, from \mbox{$150$~MHz \cite{1990MNRAS.246..256H,2017A&A...598A.104S}} to $15$~GHz \cite{2014MNRAS.438.3058R}. Its extended ($\sim$100~kpc scale) structure at low radio frequencies ($144$ and $642$~MHz) is composed of two pairs of symmetric lobes that are perpendicular to each other (i.e., roughly east--west and north--south), hinting at a sudden realignment of the jet direction in the past \cite{2022MNRAS.514.2122P}. This extended relic radio structure is rarely seen in RL-NLS1 sources \cite{2018ApJ...869..173R,2019Ap&SS.364...68G,2019MNRAS.487..640D}. To be compatible with the notion that $\gamma$-ray RL-NLS1 sources are ``young'' versions of blazars \cite{2019JApA...40...39P}, the central source in TXS~1206$+$549 may indicate episodic jet activity.  

When observed at a high resolution with very long baseline interferometry (VLBI), the source shows a compact, nearly unresolved milliarcsecond (mas)-scale structure at $5$~GHz, slightly elongated to the south  \citep{2007ApJ...658..203H,2012ApJ...744..177L}. This structure is similar to that of blazars with powerful relativistic jets, and, notably, of a couple of other $\gamma$-ray RL-NLS1 sources monitored with VLBI (see, e.g., \cite{2018rnls.confE..22L} for a review). Its accurate coordinates in the VLBI-defined International Celestial Reference Frame are right ascension $\mathrm{RA} = 12^{\mathrm{h}} 08^{\mathrm{min}} (54.256432  \pm 0.000014)^{\mathrm{s}}$ and declination $\mathrm{Dec} = 54^{\circ} 41^{\prime} (58.16329 \pm 0.00014)^{\prime\prime}$ \cite{2020A&A...644A.159C}.

In this paper, we analyse archival multi-epoch $8$-GHz VLBI data to study the jet kinematics of TXS~1206$+$549 for the first time. Tracking the positional changes in a component located in the inner $1$-mas section of the jet, identified across $5$ different epochs covering a period of $7$~yr, we determine its apparent proper motion. We also calculate core brightness temperatures, and estimate the bulk Lorentz factor and the inclination of the jet. In Section~\ref{Observations}, we describe the observational data and the analysis method. In Section~\ref{Results}, we present the results of the imaging, brightness distribution modelling, and jet parameter estimation. In Section~\ref{Discussion}, we discuss our findings and the properties of TXS~1206$+$549. Finally, a brief summary is given in Section~\ref{Conclusions}.

Throughout this paper, we assume a flat $\Lambda$ Cold Dark Matter cosmological model with Hubble constant $H_0=70$~km\,s$^{-1}$\,Mpc$^{-1}$, vacuum energy density parameter $\Omega_\Lambda = 0.7$, and matter density parameter $\Omega_\mathrm{m}= 0.3$. In this model, the luminosity distance of TXS~1206$+$549 at $z=1.344$ is $D_\mathrm{L}= 9524.4$~Mpc and the angular scale is $8.404$\,pc\,mas$^{-1}$ \cite{2006PASP..118.1711W}.

\section{Observations and Data Analysis} 
\label{Observations}

We analysed archival observations conducted with the U.S. Very Long Baseline Array (VLBA) \cite{1995ASPC...82..133W} over $6$ different epochs covering more than 24~yr between 1994 and 2018. The calibrated visibility data were acquired from the Astrogeo VLBI data base (\url{http://astrogeo.org/cgi-bin/imdb\_get\_source.csh?source=J1208\%2B5441}, accessed on 20 January 2024) maintained by L. Petrov. From the three different frequency bands (around $2$, $5$, and $8$~GHz) available in the archive, we selected $X$-band data at 8.34--8.67~GHz frequencies because their angular resolution is the highest and the time coverage is the best. This allows us to detect possible changes in the inner jet structure of TXS~1206$+$549. The short snapshot VLBA data were taken with all $10$ radio telescopes of the network with right circular polarisation. Further details (mean epochs, central observing frequencies within the $8$-GHz band, on-source integration times, bandwidths) are listed in Table~\ref{tab-obs}.

\begin{table}[H]
\caption{Details of the archival $8$-GHz VLBA observations obtained from the Astrogeo data base\label{tab-obs}}
\newcolumntype{C}{>{\centering\arraybackslash}X}
		\begin{tabularx}{\textwidth}{CCCCC}
			\toprule
			\textbf{Epoch (yr)}	& \textbf{$\boldsymbol{\nu}$ (GHz)} & \textbf{$\boldsymbol{t}$ (s)} & \textbf{IF $\boldsymbol{\times}$ BW (MHz)} & \textbf{Project} \\
			\midrule
1994.609   & 8.34	&  217 & $4\times4$ & BB023 \cite{2002ApJS..141...13B} \\
                  \midrule
2011.808   & 8.36	& 1765 & $8\times16$ & BC196 \\
                  \midrule
2014.593   & 8.67	&  109 & $12\times32$ & BG219 \cite{2016AJ....151..154G} \\
                  \midrule
2017.613   & 8.67	&  160 & $12\times32$ & UF001 \\
                  \midrule
2018.762   & 8.66	&  434 & $12\times32$ & BP222 \\
                  \midrule
2018.839   & 8.66	&  165 & $12\times32$ & UG002 \\
			\bottomrule
		\end{tabularx}
	\noindent{\footnotesize{Notes: Col.~1: mean observing epoch; Col.~2: central observing frequency; Col.~3: on-source integration time; Col.~4: number of intermediate frequency channels times bandwidth; Col.~5: project code and the literature reference (if available).}}
\end{table}

To produce radio images, we imported the calibrated visibility data into the Caltech \textsc{Difmap} program \cite{1997ASPC..125...77S}. We followed the standard hybrid mapping procedure prescribed in the \textsc{Difmap} Cookbook, with several iterations of \textsc{clean} deconvolution \cite{1974A&AS...15..417H}, phase-only, and finally phase and amplitude self-calibration, until we reached the lowest achievable noise level in the residual image. Then, we used the self-calibrated visibility data to model the source brightness distribution with two-dimensional circular Gaussian \mbox{components \cite{1995ASPC...82..267P}}. The parameters of these model components describe the source structure.

\section{Results} 
\label{Results}

\subsection{Source Structure}

\begin{figure}[H]
\includegraphics[width=0.5\columnwidth]{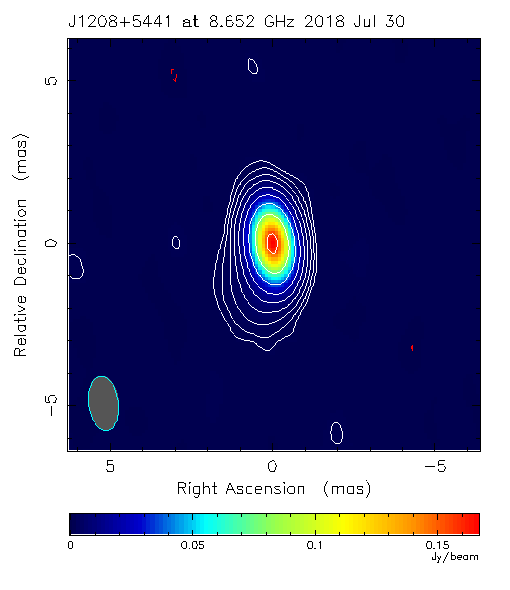}
\includegraphics[width=0.5\columnwidth]{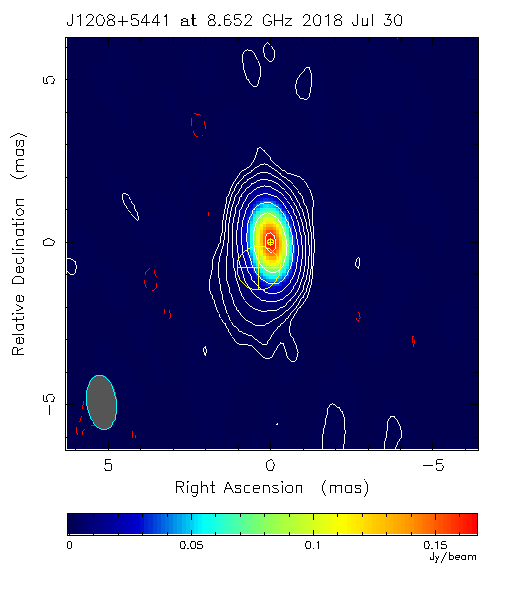}
\caption{Naturally weighted 8.7-GHz VLBA images of TXS~1206$+$549 from 30 July 2018, deconvolved with \textsc{clean} components {(\textbf{left})} and a model composed of two Gaussian components whose positions and FWHM diameters are indicated with cross-hairs in yellow circles {(\textbf{right})}. In both images, the peak intensity is $167$~ mJy\,beam$^{-1}$, the contour levels start at $\pm 0.6$~ mJy\,beam$^{-1}$ ($\sim$$3\sigma$ image noise), and further positive contours increase by a factor of 2. The elliptical Gaussian restoring beam size is $1.69\,\mathrm{mas} \times 0.92\,\mathrm{mas}$ (FWHM), the major axis position angle is $7.2^{\circ}$ (measured from north through east), as indicated in the bottom left corner of the images. \label{fig-map}}
\end{figure}

The mas-scale radio structure of TXS~1206$+$549 is compact at $8$~GHz, as indicated in Figure~\ref{fig-map}, shown as an example. In all six epochs, from 1994 to 2018, the brightness distribution modelling reveals a compact core (C), which corresponds to the synchrotron, self-absorbed base of the relativistic jet, and a more extended and weaker jet component can be seen south--southeast of the core. The parameters of the fitted circular Gaussian model components are listed in Table~\ref{tab-fit}. The sizes of the components always exceeded the minimum resolvable angular size with the interferometer array \cite{2005AJ....130.2473K}. We generally adopted the method for parameter error estimates from \cite{2008AJ....136..159L}. However, the formal errors of flux density values were conservatively set at $10\%$ to account for the possible VLBI absolute amplitude calibration uncertainties \cite{2021ApJ...919...40P}. The uncertainty of the jet component position angle is taken as $1^{\circ}$. For an error estimation of the size of the fitted Gaussian components, we used the approach \mbox{in \cite{2014MNRAS.445.1370K}}, which provides more realistic errors for fitted components that are smaller than the restoring beam size.   

\vspace{-3pt}

In the first epoch, the core--jet component separation is $1.56$~mas, while in later epochs, starting from 2011, the separation is within $1$~mas. We interpret these as two different components in the jet structure of TXS~1206$+$549, denoted by J1 (in 1994) and J2 (from 2011 to 2018). Apparently the J1 component moved away from the core and became too faint and extended for detection with VLBI by the 2010s. During the $17$ yr long gap in the archival observations, a new component, J2, must have emerged, apparently following the track of the earlier J1. Since the time sampling of the data in the 2010s is much better, we could use the last five epochs between 2011 and 2018 for the jet kinematic study of J2.

\begin{table} [H]
\caption{Brightness distribution model parameters and brightness temperatures derived for TXS~1206+549 from $8$-GHz VLBA imaging data\label{tab-fit}}
\newcolumntype{C}{>{\centering\arraybackslash}X}
	\begin{adjustwidth}{-\extralength}{0cm}
\begin{tabularx}{\fulllength}{CCCCCCC}
\toprule
			\textbf{Epoch (yr)}	& \textbf{Comp.} & \textbf{S (mJy)}	& \textbf{R (mas)}     & \textbf{$\boldsymbol{\phi}$ ($^{\boldsymbol{\circ}}$)} 	& \textbf{d (mas)} & \textbf{T$_{\boldsymbol{\mathrm{b}}}$ ($\boldsymbol{10}^{\boldsymbol{10}}$~K)}\\
			\midrule
\multirow[m]{2}{*}{1994.609}	& C			&  228 $\pm$ 23		& $\dots$ & $\dots$	& $0.38 \pm 0.01$ & 6.5 $\pm$ 0.4\\
			  	                
			             	      & J1		& 30   $\pm$ 3    & 1.56 $\pm$ 0.03   & 131 & $1.45 \pm 0.09$ & $\dots$ \\
                   \midrule
\multirow[m]{2}{*}{2011.808}    & C			& 233 $\pm$ 23		& $\dots$ & $\dots$	& $0.309 \pm 0.004$ & 10.0 $\pm$ 0.4\\
			  	            
			             	     & J2			& 43   $\pm$ 4    & 0.49 $\pm$ 0.04     & 150 & $1.93 \pm 0.03$ & $\dots$ \\
                   \midrule
\multirow[m]{2}{*}{2014.593}    & C			& 119 $\pm$ 12		& $\dots$ & $\dots$	& $0.365  \pm  0.005$ & 3.4 $\pm$ 0.2\\
			  	            
			             	    & J2			& 20  $\pm$ 2   & 0.73 $\pm$ 0.04       & 163 & $1.35  \pm  0.04$ & $\dots$ \\
                  \midrule
\multirow[m]{2}{*}{2017.613}   & C			& 307  $\pm$ 31		& $\dots$ & $\dots$	& $0.144  \pm  0.002$ & 56 $\pm$ 2\\
			  	               
			             	    & J2			& 19    $\pm$ 2   & 0.83 $\pm$ 0.03      & 157 & $0.83 \pm  0.03$ & $\dots$ \\
                     \midrule
\multirow[m]{2}{*}{2018.762}   & C			& 167 $\pm$ 17  		& $\dots$ & $\dots$	& $0.191  \pm  0.003$ & 17.5 $\pm$ 0.6 \\
			  	               
			             	    & J2		& 18  $\pm$ 2     & 0.90 $\pm$ 0.03       & 155 & $1.40  \pm  0.03$ & $\dots$ \\
                     \midrule
\multirow[m]{2}{*}{2018.839}   & C		& 181 $\pm$ 18		& $\dots$ & $\dots$	& $0.289   \pm 0.006$ & 8.3 $\pm$ 0.4\\
			  	               
			             	    & J2			                                              & 19   $\pm$ 2    & 0.91 $\pm$ 0.03      & 148 & $2.86   \pm 0.06$ & $\dots$ \\
 \bottomrule
\end{tabularx}
\end{adjustwidth}
 \noindent{\footnotesize{Notes: Col.~1: mean observing epoch; Col.~2: component designation; Col.~3: flux density; Col.~4: radial angular distance from the core; Col.~5: position angle with respect to the core, measured from north through east; Col.~6: circular Gaussian component diameter; Col.~7: core brightness temperature.}}
\end{table}
\vspace{-15pt}

\subsection{Jet Component Proper Motion}

\begin{figure}[H]
\includegraphics[width=10.5 cm]{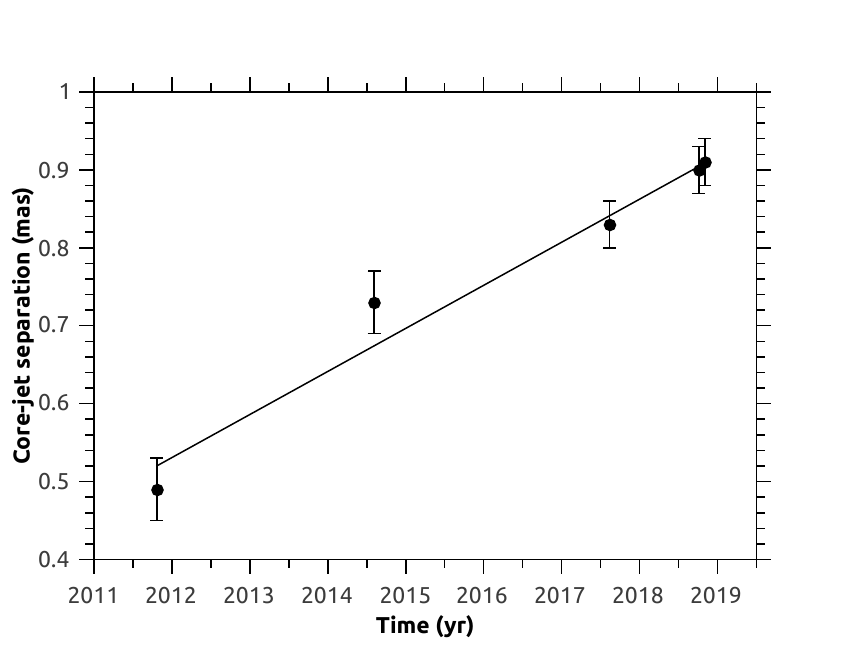}
\caption{The core--jet separation for the J2 component in TXS~1206$+$549 as a function of time, based on VLBI observations and model fitting at $8$~GHz in five epochs covering a period of about $7$~yr. Linear regression indicates $\mu = (0.055 \pm 0.006)$~mas\,yr$^{-1}$ apparent proper motion. \label{fig-propm}}
\end{figure}

Based on our fitted brightness distribution models (Table~\ref{tab-fit}), the angular separation between the core and J2 components monotonically increased from 2011 to 2018, consistently with an outward jet motion. The values are plotted in Figure~\ref{fig-propm}. Assuming there is a linear change in the core--jet separation, the apparent proper motion of J2 is $\mu = (0.055 \pm 0.006)$~mas\,yr$^{-1}$. Extrapolating back in time, the component ejection could have occurred around the year~2002.

\vspace{-27pt}

This proper motion can be expressed as the speed of light ($c$) as follows:
\begin{equation}
    \beta_\mathrm{app} = 0.0158 \, \frac{\mu D_\mathrm{L}}{1+z},
\end{equation}
where $\mu$ is measured in mas\,yr$^{-1}$ and the luminosity distance $D_\mathrm{L}$ is expressed in Mpc, e.g.,~\cite{2012ApJ...760...77A}. By substituting $z=1.344$ and $D_\mathrm{L}=9524.4$~Mpc, we can obtain $\beta_\mathrm{app} = 3.5 \pm 0.4$ apparent transverse speed. As $\beta_\mathrm{app}$ clearly exceeds unity, the J2 component is characterised by apparent superluminal motion.

\subsection{Jet Parameters}

The knowledge of the apparent transverse speed in the inner jet, together with the measurement of the brightness temperature in the compact core, allow us to estimate the basic physical and geometric parameters of the relativistic jet in TXS~1206$+$549.

The core brightness temperature values listed in the last column of Table~\ref{tab-fit} for each observed epoch are calculated as follows: \cite{1982ApJ...252..102C}
\begin{equation} \label{eq:tb}
    T_{\rm{b}} = 1.22 \times 10^{12} \, (1 + z) \frac{S}{d^2 \nu^2} \,\, \mathrm{K}.
\end{equation}
Here, the core flux density $S$ is measured in Jy, the observing frequency $\nu$ in GHz, and the fitted circular Gaussian component diameter (FWHM) $d$ in mas. The measured brightness temperatures are close to or exceed the equipartition value, $T_\mathrm{b,eq} \approx 5 \times 10^{10}$\,K \citep{1994ApJ...426...51R}. The latter indicate Doppler-boosted radio emissions from the jet. Since brightness temperatures vary considerably (by more than an order of magnitude) from epoch to epoch, we followed the approach of \cite{2021ApJ...923...67H} and took the median value, $T_\mathrm{b,med} = 9.1 \times 10^{10}\,\mathrm{K}$ as characteristic to the jet. The advantage of using the median brightness temperature is that this value is insensitive to outliers due to source variability. Note that by far the highest measured brightness temperature, $T_\mathrm{b} = (5.6 \pm 0.2) \times 10^{11}\,\mathrm{K}$, is caused by a major outburst, as indicated by the high core flux density in 2017. According to Equation~(\ref{eq:tb}), the small angular size of the core component also contributes to the high $T_\mathrm{b}$ (Table~\ref{tab-fit}).   

The characteristic Doppler-boosting factor is calculated as  follows:
\begin{equation} \label{eq:delta}
 \delta = \frac{T_\mathrm{b,med}}{T_\mathrm{b,int}},
\end{equation}
where we adopt $T_\mathrm{b,int} = 4.1 \times 10^{10}\,\mathrm{K}$ as the intrinsic brightness temperature.  This is a typical value found for a sample of radio quasars that are not in the outburst \cite{2021ApJ...923...67H}. Notably, $T_\mathrm{b,int} \lesssim T_\mathrm{b,eq}$. For the jet in TXS~1206$+$549, we obtained $\delta=2.2$.

From the characteristic Doppler-boosting factor and the apparent transverse speed of the inner sub-mas jet component J2, the bulk Lorentz factor ($\Gamma$) of the plasma and the viewing angle of the jet ($\vartheta$) with respect to the line of sight can be calculated, e.g., \cite{1995PASP..107..803U}:
\begin{equation} \label{eq:Gamma}
    \Gamma = \frac{\beta_\mathrm{app}^2 + \delta^2 + 1}{2\delta},
\end{equation}
\begin{equation} \label{eq:theta}
    \vartheta = \arctan \left({\frac{2\beta_\mathrm{app}}{\beta_\mathrm{app}^2 + \delta^2 - 1}}\right).
\end{equation}
After substituting the numerical parameters derived above, $\beta_\mathrm{app} = 3.5 \pm 0.4$ and $\delta=2.2$, we obtain $\Gamma= 4.1 \pm 0.6$ and $\vartheta=(23.5 \pm 0.9)^{\circ}$. 

To investigate the uncertainties inherent in our method of using $T_\mathrm{b,med}$ for estimating the characteristic Doppler factor (Equation~(\ref{eq:delta})), we calculated possible lower and upper limits for the Doppler factor, based on individual $T_\mathrm{b}$ measurements taken from all observing epochs but the one in 2017 (Table~\ref{tab-fit}). The exclusion of the 2017 epoch when the source was in outburst is justified because equipartition conditions are known to be violated in the maximum brightness state \cite{2006ApJ...642L.115H}. Therefore, using the intrinsic brightness temperature characteristic of the low state in the denominator of Equation~(\ref{eq:delta}) would lead to an unrealistically high Doppler factor value. Assuming the range $3.4 \times 10^{10}~\mathrm{K} \le T_\mathrm{b} \le 1.75 \times 10^{11}~\mathrm{K}$, and again adopting $T_\mathrm{b,int} = 4.1 \times 10^{10}\,\mathrm{K}$ as the intrinsic brightness temperature \cite{2021ApJ...923...67H}, the allowed range of Doppler factors is $ 0.8 \le \delta \le 4.3$. While the lower limit implies no Doppler boosting, the upper limit, together with the apparent superluminal speed found for the jet component J2, would result in a $3.7$ Lorentz factor and $13.2^{\circ}$ jet inclination (Equations~(\ref{eq:Gamma}) and (\ref{eq:theta})).

\section{Discussion} 
\label{Discussion}

According to our analysis of $8$-GHz VLBA data, the jet in TXS~1206$+$549 is characterised by a moderate Lorentz factor, $\Gamma \approx 4$. Studies of powerful jetted AGN samples \mbox{e.g., \cite{2019ApJ...874...43L}} found that the distribution of Lorentz factors peaks between $5$ and $15$, with high values occasionally reaching $\Gamma_\mathrm{max} \approx 50$. The inclination of the jet ($\vartheta \approx 24^{\circ}$) is larger than expected from a blazar-type source, where the jet is supposed to closely align with the line of sight within about $10^{\circ}$. Note, however, that even a moderate jet bending well inside the region resolvable with VLBI at $8$~GHz (i.e., within <1~mas, corresponding to <8~pc in projected linear size) might cause the innermost section of the jet to be better aligned to the line of sight. This could possibly be tested with higher-resolution VLBI imaging at higher frequencies. Nevertheless, the jet inclination is generally consistent with the Sy1 classification of the galaxy (i.e., seen face-on) if we naturally assume that the jet direction is perpendicular to the plane of the accretion disk and the obscuring ``torus''. 

With the apparent speed measured for the jet component J2 ($\beta_\mathrm{app} = 3.5$), a Doppler factor of $\delta \gtrsim 5.3$ would be required for a blazar-like jet inclination. Note that we characterised the jet using the median core brightness temperature $T_\mathrm{b,med}$, and there were individual $T_\mathrm{b}$ measurements that could, in principle, lead to higher Doppler factor estimates according to Equation~(\ref{eq:delta}). Indeed, the range of brightness temperatures measured at epochs when the source was in a low state would allow for a $\delta$ of as high as $4.3$ and a jet inclination as small as $13.2^{\circ}$. However, the extremely high $T_\mathrm{b}=(5.6 \pm 0.2) \times 10^{11}$~K measured in 2017 is clearly associated with an outburst when the intrinsic brightness temperature can significantly exceed the equipartition value, with characteristic value estimated as \mbox{$T_\mathrm{b,int} \sim 2 \times 10^{11}$~K \cite{2006ApJ...642L.115H}}. If so, the Doppler factor is comparable to the value of $\delta=2.2$ we determined from the median brightness temperature that represents the quiescent state of the jet. Therefore, we believe our characteristic Doppler factor estimate is robust.

TXS~1206$+$549 is known to have rapid, high-amplitude flux density variations in the radio. It is part of the AGN sample monitored at $15$~GHz with the Owens Valley Radio Observatory (OVRO) 40-m radio telescope \cite{2011ApJS..194...29R}. For example, the published light curve \cite{2014MNRAS.438.3058R} (available at \url{https://vizier.cds.unistra.fr/viz-bin/VizieR?-source=J/MNRAS/438/3058}, accessed on 20 January 2024) indicates almost a factor of $4$ change in total flux density during the relatively short interval between April 2009 and December 2011, with a typical peak-to-peak time difference of just over $\sim$$1$~yr. Unfortunately, this monitoring interval has little overlap with our archival VLBA data (Table~\ref{tab-obs}). Close to our only epoch falling in the available OVRO monitoring time range, $2011.808$~yr, the OVRO $15$-GHz flux density was at around its local minimum, with $\sim$$130$~mJy. With an OVRO light curve continuing into the 2010s, it would be possible to independently verify the outburst seen in 2017 in our sparsely time-sampled VLBA data.

The OVRO flux density observations were applied to estimate variability brightness temperatures for the sources in the monitored sample, including TXS~1206$+$549 \cite{2018ApJ...866..137L}. The method is based on fitting flares identified in the radio light curve with exponential rise and decay functions. The value $T_\mathrm{var} = 5 \times 10^{14}$~K found for TXS~1206$+$549 led to the estimate of the variability Doppler factor $\delta_\mathrm{var} = 29.09^{+10.28}_{-10.02}$. This is significantly, by an order of magnitude, higher than our estimate based on the measured median brightness temperature and the assumed intrinsic brightness temperature, which is close to the equipartition value. A comparison of TXS~1206$+$549 with other $\gamma$-ray-emitting RL-NLS1 sources indicates that the characteristic brightness temperature we determined is similar to those found for other sources, in the order of $T_\mathrm{b} \sim 10^{10-11}$~K \cite{2015MNRAS.453.4037O,2019MNRAS.487L..40Y,2012ApJ...744..177L,2023ApJ...943..136S}. 

Under the name of the \textit{Fermi} $\gamma$-ray source 4FGL~J1208.9$+$5441, our RL-NLS1 object has various estimates of the Doppler factor found in the literature. The value of 15.24 was determined for the so-called luminosity Doppler factor derived from the observed $\gamma$-ray luminosity and the photon index \cite{2020AN....341..462X}. Another publication dealing with \textit{Fermi} blazar jet \mbox{properties \cite{2022ApJ...925...40X}} arrived at $\delta>8.12$ by comparing the jet radiation power and the $\gamma$-ray luminosity, assuming that the $\gamma$-ray luminosity is representative of the jet bolometric luminosity and $\delta=\Gamma$ for blazars. (The latter assumption does not hold according to our VLBI results.) This paper  \cite{2022ApJ...925...40X} also provides an alternative value of 15.12 calculated based on a correlation between the jet radiative power and the luminosity of the BLR \cite{2020ApJ...897...10Z}. However, another method, based on a fitting model for the broad-band spectral energy distribution (SED), again assuming equality between the Doppler factor and the Lorentz factor, led to a different value: $\delta=7.8$ \cite{2018ApJS..235...39C}. Most recently, $\delta=25.69$ was derived for 4FGL~J1208.9$+$5441, using SED fitting to quasi-simultaneous multi-wavelength data and assuming a one-zone leptonic model \cite{2023ApJS..268....6C}.

As we can see from the values collected above, a wide range of Doppler factors can be estimated for TXS~1206$+$549, with our $\delta=2.2$ being the lowest. Clearly, there are several differences in the  methods used in the literature, for example, in the frequency bands that are observed, and therefore the regions of the jet that are probed, in the observed epoch (i.e., in quiescence or outburst), and in the physical model assumptions used. It is known that while different determinations of the Doppler factors are generally statistically consistent with each other for populations of sources, there might indeed be large discrepancies for individual objects \cite{2017A&A...602A.104L}.  

The cited studies estimating the Doppler factor from high-energy data consider TXS~1206$+$549 as a blazar. However, there are no generally accepted criteria for the blazar classification of AGN. The working definition of blazars could vary from waveband to waveband, depending of the types of observing data that are available and the purpose of the study. The basic concept is that blazars are jetted AGN, whose relativistic jet (more specifically, one of the intrinsically symmetric bipolar jets on opposite sides of the central BH) is directed very closely to the observer. Exactly how closely is a matter of definition. The $\vartheta < 10^{\circ}$ criterion, e.g., \cite{1999ApJ...521..493L,2021ApJ...909..103Z,2022Galax..10...35P}, that we adopt in this paper is probably the simplest, although some authors cite $\vartheta < 20^{\circ}$, e.g., \cite{2024arXiv240104009X}. Alternatively, for statistical studies, it can be practical to apply $\vartheta \le 1/\Gamma$, e.g., \cite{2015MNRAS.446.2483S,2021Galax...9...23S}. Note that, according to this latter definition and based on our Lorentz factor estimate, $\Gamma = 4.1$, the jet inclination angle upper limit in our case would be $\vartheta_\mathrm{lim} = 0.2439\,\mathrm{rad} \approx 14^{\circ}$, still lower than the $\vartheta \approx 24^{\circ}$ we derived. In any case, the jet inclination angle can only be measured indirectly from VLBI kinematic studies similar to ours  (Section~\ref{Results}) and using certain assumptions for the jet physics. Moreover, in large part because such measurements require considerable observational resources at multiple epochs, this method is only applicable to a very limited sample of the sources compared to all blazar candidates. Thus, viewing angles are not too practical to use for a definition.

As a consequence of the closely aligned relativistic jets with respect to the line of sight in blazars, VLBI observations often indicate the apparent superluminal motion of the jet components. However, on the one hand, very small inclinations ($\vartheta \approx 0^{\circ}$) do not lead to detectable transverse motion in the plane of the sky. On the other hand, $\beta_\mathrm{app} > 1$ values are possible even at relatively high jet inclinations (see, e.g., Figure~21 in \cite{1995PASP..107..803U}). According to our results, this is the case for TXS~1206$+$549 as well. Therefore, the detection of superluminal motion cannot be used as a defining criterion for blazars. 

Small-inclination relativistic jets in AGN produce highly enhanced synchrotron radiation via Doppler boosting, high-energy emissions via the inverse Compton process, and may show rapid and high-amplitude variability in the broadest possible range of electromagnetic wavebands. Therefore, these criteria are often applied for a ``loose'' definition of blazars. For example, the important and widely used multi-frequency catalogue of blazars, the Roma-BZCAT \cite{2009A&A...495..691M,2015Ap&SS.357...75M} collection contains AGN with optical identification and blazar-like spectrum, high X-ray luminosity, and radio detection at $\sim$$1$~GHz frequency with ``compact'' radio morphology. However, the radio emission confined to within $\sim$$10^{\prime\prime}$ angular scales does not guarantee the kind of compactness one would expect from a closely-aligned jet. Indeed, it was found recently that a significant fraction, $3\%$ of the so-called confirmed Roma-BZCAT entries show an extended radio structure at $\sim$$1^{\prime\prime}$ scale and are likely misclassified as blazars \cite{2024arXiv240104009X}, not to mention VLBI imaging, which provides a higher resolution by orders of magnitude to potentially refute the blazar nature of certain candidates, e.g., \cite{2017MNRAS.467..950C}. Finally, to demonstrate the confusion about the blazar classification in the literature, we shall briefly mention the so-called ``radio-weak'' BL Lac objects, e.g., \cite{2017ApJ...834..113M,2018ApJ...854L..23B,2019MNRAS.482L..34C}. The term itself is contradictory since the presence of a small-inclination relativistic jet is an essential ingredient of a blazar, and should be a source of Doppler-beamed radio emission via the synchrotron process.  

As was found from multi-waveband studies of a large sample of $42$ flat-spectrum RL-NLS1 sources \cite{2015A&A...575A..13F}, their central BH masses are lower than those for blazars, while the accretion rates are generally comparable. The jet power per unit mass is very similar for these different types of jetted AGN, suggesting that the observational differences are mainly caused by scaling effects. In other words, the jet launching mechanism in the RL-NLS1 sources is similar to that in blazars, e.g., \cite{2019Ap&SS.364..123C}, the only major difference being the BH mass, which is lower in RL-NLS1 sources. For a sample of $16$ $\gamma$-ray RL-NLS1 sources, it was found that the jet power in almost all of them is consistent with a hybrid jet formation model involving a mixture of both the Blandford--Payne and Blandford--Znajek mechanisms \cite{2022MNRAS.517.1381C}. If the general picture regarding the similar jet launching required for blazars and RL-NLS1 sources is true, then there is no obvious evolutionary connection between RL-NLS1 sources and FSRQs that harbour one or two orders of magnitude more massive central BHs. Because of the low-mass BH, variability time scales in RL-NLS1 sources are shorter than those expected from Doppler boosting only \cite{2015A&A...575A..13F}. In the case of TXS~1206$+$549, this might explain the much higher variability Doppler factor compared to our jet kinematic estimate. 

Concerning the dichotomy between the radio-loud and radio-quiet AGN, the properties of the SMBHs, as well as their environments, should play a role. Simulations suggest that the jet power is a strong function of the BH spin in the presence of a thick accretion disk \cite{2010ApJ...711...50T}. However, observations indicate that most SMBHs rotate rapidly \cite{2002ApJ...565L..75E}; therefore, a high black hole spin alone cannot be responsible for the jet production, and disk magnetization could also be an important factor (see the discussion and references \mbox{in \cite{2019ARA&A..57..467B}}). In any case, as the radio-loud fraction of NLS1 sources is similar to that of the general AGN population, it is reasonable to assume that the same effects govern these objects, which have lower-mass BHs in their central engine.

\section{Summary and Conclusions} 
\label{Conclusions}

We analyzed the mas-scale radio structure of the $\gamma$-ray-detected RL-NLS1 galaxy TXS~1206$+$549 at $z=1.344$. This is the highest-redshift $\gamma$-ray NLS1 known to date \cite{2021MNRAS.504L..22R}. We used the archival VLBA data available in the Astrogeo data base in $6$ epochs from $1994$ to $2018$. Imaging and brightness distribution modelling allowed for us to track the motion of a jet component within $1$~mas from the core from $2011$ to $2018$, and to determine its superluminal apparent transverse speed $\beta_\mathrm{app} = 3.5 \pm 0.4$. We characterised the brightness temperature of the compact core with the median of the measured values, and derived the Doppler factor $\delta=2.2$ by assuming an intrinsic brightness temperature typical for jetted AGN that are not in a flux density outburst. From the apparent transverse speed and the characteristic Doppler factor, we obtained $\Gamma= 4.1 \pm 0.6$ for the bulk Lorentz factor for the jet plasma and $\vartheta=(23.5 \pm 0.9)^{\circ}$ for the jet inclination angle with respect to the line of sight. The latter is consistent with the Sy1 classification of the source.

Some properties of TXS~1206$+$549---the $\gamma$-ray emission \cite{2021MNRAS.504L..22R}, the superluminal jet motion, and the rapid flux density variability in multiple wavebands \cite{2021MNRAS.504L..22R} in particular---are similar to those of powerful blazars. Because the jet inclination angle is found to exceed (10--20)$^{\circ}$, strictly speaking, we did not find strong evidence for its blazar-like nature from the $8$-GHz VLBI imaging data. However, blazar definitions are usually more liberal in the literature, and $\vartheta$ is not particularly large. A moderate bending of the innermost jet section could help to reconcile TXS~1206$+$549 with a blazar-type object in terms of jet inclination angle. Dedicated higher-resolution VLBI imaging at higher observing frequencies would be beneficial to test this possibility and probably also to constrain the core brightness temperature more tightly.

\vspace{6pt}
\authorcontributions{Conceptualization, S.F.; methodology, B.K. and S.F.; validation, S.F. and K.\'E.G.; formal analysis, B.K. and S.F.; visualization, B.K. and S.F.; writing---original draft preparation, B.K.; writing---review and editing, B.K., S.F. and K.\'E.G.; supervision, S.F. All authors have read and agreed to the published version of the manuscript.}

\funding{This research was funded by the Hungarian National Research, Development and Innovation Office (NKFIH), grant number OTKA K134213. This project has received funding from the HUN-REN Hungarian Research Network.}

\dataavailability{The calibrated VLBI data are available from the Astrogeo archive (\url{http://astrogeo.org/} (accessed on 20 January 2024) and \url{https://astrogeo.smce.nasa.gov/vlbi_images/} (accessed~on 20 January 2024).} 

\acknowledgments{We thank the anonymous referees for their comments and suggestions. The National Radio Astronomy Observatory is a facility of the National Science Foundation operated under cooperative agreement by Associated Universities, Inc.
We acknowledge the use of archival calibrated VLBI data from the Astrogeo Center database maintained by Leonid Petrov.}

\conflictsofinterest{The authors declare no conflicts of interest.} 

\abbreviations{Abbreviations}{
The following abbreviations are used in this manuscript:\\

\noindent 
\begin{tabular}{@{}ll}
AGN & active galactic nuclei\\
BH & black hole\\
FSRQ & flat-spectrum radio quasar\\
FWHM & full width at half maximum\\
NLS1 & narrow-line Seyfert 1 galaxy\\
NRAO & National Radio Astronomy Observatory\\
QPO  & quasiperiodic oscillation\\
RL-NLS1 & radio-loud narrow-line Seyfert 1 galaxy\\
SED & spectral energy distribution\\
SMBH & supermassive black hole\\
SNR & signal-to-noise ratio\\
Sy1 & Seyfert 1 galaxy\\
Sy2 & Seyfert 2 galaxy\\
VLBA & Very Long Baseline Array\\
VLBI & very long baseline interferometry\\
\end{tabular}
}

\begin{adjustwidth}{-\extralength}{0cm}

\reftitle{References}



\PublishersNote{}
\end{adjustwidth}
\end{document}